%% file: main.tex
\documentclass[10pt, conference]{IEEEtran}
\IEEEoverridecommandlockouts
\usepackage{cite}
\usepackage{amsmath,amssymb,amsfonts}
\usepackage{algorithmic}
\usepackage{graphicx}
\usepackage{textcomp}
\usepackage{xcolor}
\usepackage[linesnumbered,ruled,vlined]{algorithm2e}
\usepackage{verbatim}
\usepackage{makecell}
\usepackage{graphicx}
\usepackage{float}
\usepackage{subfig}
\usepackage{url}
\usepackage{setspace}
\usepackage{tabularx}

\def\BibTeX{{\rm B\kern-.05em{\sc i\kern-.025em b}\kern-.08em
    T\kern-.1667em\lower.7ex\hbox{E}\kern-.125emX}}

\begin{document}
\bstctlcite{IEEEexample:BSTcontrol}

\title{TI-DNS: A Trusted and Incentive DNS Resolution Architecture based on Blockchain
}

\author{
        \IEEEauthorblockN{Yufan Fu\IEEEauthorrefmark{1}\IEEEauthorrefmark{2}, Jiuqi Wei\IEEEauthorrefmark{1}\IEEEauthorrefmark{2}, Ying Li\IEEEauthorrefmark{1}\IEEEauthorrefmark{2}, Botao Peng\IEEEauthorrefmark{1}, Xiaodong Li\IEEEauthorrefmark{1}\thanks{Xiaodong Li is the corresponding author.}}

		\IEEEauthorblockA{
		\IEEEauthorrefmark{1}Institute of Computing Technology, Chinese Academy of Sciences\\
		}
		
		\IEEEauthorblockA{
		\IEEEauthorrefmark{2}University of Chinese Academy of Sciences\\
		}
		
		\IEEEauthorblockA{\{fuyufan20z, weijiuqi19z, liying18z, pengbotao, xl\}@ict.ac.cn}
}

\maketitle

\begin{abstract}
Domain Name System (DNS) is a critical component of the Internet infrastructure, responsible for translating domain names into IP addresses. However, DNS is vulnerable to some malicious attacks, including DNS cache poisoning, which redirects users to malicious websites displaying offensive or illegal content. Existing countermeasures often suffer from at least one of the following weakness: weak attack resistance, high overhead, or complex implementation. To address these challenges, this paper presents TI-DNS, a blockchain-based DNS resolution architecture designed to detect and correct the forged DNS records caused by the cache poisoning attacks in the DNS resolution process. TI-DNS leverages a multi-resolver \textit{Query Vote} mechanism to ensure the credibility of verified records on the blockchain ledger and a stake-based incentive mechanism to promote well-behaved participation. Importantly,  TI-DNS is easy to be adopted as it only requires modifications to the resolver side of current DNS infrastructure. Finally, we develop a prototype and evaluate it against alternative solutions. The result demonstrates that TI-DNS effectively and efficiently solves DNS cache poisoning.

\end{abstract}


\input{intro}

\input{sys}




\bibliographystyle{IEEEtran}
\bibliography{IEEEabrv, refbib}

\end{document}

%% file: intro.tex
\vspace{-0.2em}

\section{Introduction}

The Domain Name System (DNS) is a  distributed naming system that enables the translation of human-readable domain names into IP addresses\cite{rfc1034}. DNS plays a crucial role in ensuring the reliability of Internet services, allowing users to browse websites, send emails, and perform various other network functions. However, the distributed nature of DNS and its reliance on multiple parties to provide accurate information make it vulnerable to various security attacks.
One particular security threat facing DNS is cache poisoning attack\cite{klein2017internet}, where several methods exist to inject false information into a DNS resolver by exploiting protocol or software vulnerabilities\cite{klein2007openbsd, duan2012hold, zhang2021study}. Once the cache of a resolver has been poisoned, any device dependent on that resolver will be directed to the attacker's website instead of the legitimate one. Therefore, successful cache poisoning attacks can have severe consequences, such as phishing, malware injection, and Denial of Service\cite{DNSopearating2022}.

To combat the threat of cache poisoning attacks, Source Port Randomization (SPR)\cite{rfc5452} was proposed and widely adopted by many software vendors. SPR employs the idea of increasing entropy. Instead of using a fixed or sequential port number, the resolver selects a random port from the available range for each outstanding query. The attackers must accurately guess both the transaction ID and the random source port in a DNS query. There 
are also other efforts made to further increase the entropy\cite{dagon2008increased, perdisci2009wsec, herzberg2013fragmentation}. However, all these mechanisms can only take effect once all DNS servers fully support them, which restricts their practical adoption\cite{zhang2021study}. Moreover, they are vulnerable to adversaries who use Man-in-the-Middle (MitM) mechanism to intercept communication\cite{DNSopearating2022}.

Another widely adopted approach is the deployment of DNSSEC (DNS Security Extensions)\cite{rfc6840}, which provides a cryptographic mechanism to verify the authenticity and integrity of DNS responses. This solution has made significant progress in improving the security of DNS and mitigates the risk of cache poisoning. However, the deployment and implementation of DNSSEC are going extremely slowly\cite{APDNSSEC}. Besides, DNSSEC introduces additional complexity and overhead to DNS interactive architecture and processes, which impacts the overall performance and efficiency of DNS resolution\cite{lu2019end}.

Some other redundant architectures like HARD-DNS\cite{gutierrez2010hard}, DepenDNS\cite{alfardan2011analysis} and CGuard\cite{chau2018adaptive} require making some changes to the current DNS infrastructure. Although these modifications partially address DNS security issues like cache poisoning, their practical application is difficult due to their relatively high deployment costs and performance losses.  

\textbf{[Our Solution]} Based on these observations, we propose TI-DNS, a trusted and incentive DNS resolution architecture based on blockchain, which can detect and correct forged DNS records caused by cache poisoning attacks. Compared to existing solutions, TI-DNS offers enhanced anti-attack capabilities with similar or even reduced overhead. Moreover, its implementation is relatively simple, only requiring modifications on the resolver side of the current DNS. In summary, TI-DNS has the following characteristics:

 \textbf{Trusted.} Each TI-DNS network maintains a globally unique blockchain ledger as a \textit{Verification Cache} for resolvers. If the response from the authoritative server is forged, it must differ from the one indicated by the \textit{Verification Cache}. Besides, to ensure the ledger's credibility, we design a multi-resolver \textit{Query Vote} mechanism implemented by the smart contract to validate each record create or update operation. Compared to other redundant architectures similar to ours\cite{gutierrez2010hard, sun2009dependns, yu2020dnstsm}, TI-DNS shows a stronger ability to resisit attacks (enhanced by 1-3 orders of magnitude), and is also more resistant to internal attacks.

\textbf{Incentive.} To promote participation, TI-DNS follows the convention of some decentralized applications where stake rewards are presented to the participants (i.e., resolvers) that behave well. In the \textit{Query Vote} process, we design a \textit{Voter Selection} algorithm that chooses some validators according to the proportion of stake they hold, which means participants with better historical behavior have greater decision-making power, thus further enhancing the reliability of the system. 

\textbf{Efficient.} TI-DNS adds the process of verifying the authoritative response by querying the \textit{Verification Cache}. However, TI-DNS outperforms the cryptographic-based method DNSSEC (about 5\% faster) since it does not require any encryption/decryption process. Additionally, in comparison to other similar redundant architectures, TI-DNS outperforms theirs (35\% $\sim$ 65\% faster) because it does not rely on multiple resolvers to provide the response in each resolution process. 
    
\textbf{Practical.} Compared to some existing schemes, we believe TI-DNS is easier to be adopted: (1) TI-DNS only requires modifications to the resolver side of the existing DNS infrastructure (unlike DNSSEC must modify all levels), and does not need to design new interaction protocols or introduce new types of resource records. (2) TI-DNS has good backward compatibility, that is, it can be compatible with existing DNS extension protocols, like DNSSEC, DoT\cite{rfc7858}, DoH\cite{rfc8484}, etc.

Our contributions are summarized as follows:

\begin{itemize}
    \item We propose TI-DNS, a blockchain-based DNS resolution architecture, to solve the problem of cache poisoning. 
    
    \item We design a three-phase \textit{Query Vote} procedure to manage the verified records in the blockchain ledger, which ensures the ledger's credibility as well as the transparency and traceability of participants' operations.
    
    \item We design a stake-based \textit{Voter Selection} and \textit{Incentive} mechanism, which can evaluate the credibility of various resolvers and encourage well-behaved participation. To our knowledge, TI-DNS is the first to introduce incentives in solving the DNS cache poisoning.
    
    \item We develop a prototype system based on SDNS\cite{SDNS} and Hyperledger Fabric\cite{androulaki2018hyperledger}, and prove its effectiveness and efficiency through experiments. The result demonstrates that TI-DNS outperforms existing solutions in terms of both attack resistance and resolution performance.

\end{itemize}



\section{Background and Related Work}
\label{Sec.Related Work}

\subsection{Kaminsky DNS Cache Poisoning}
\label{sec-kaminsky}

\begin{figure}[tb]
\centering
\includegraphics[width=1\columnwidth]{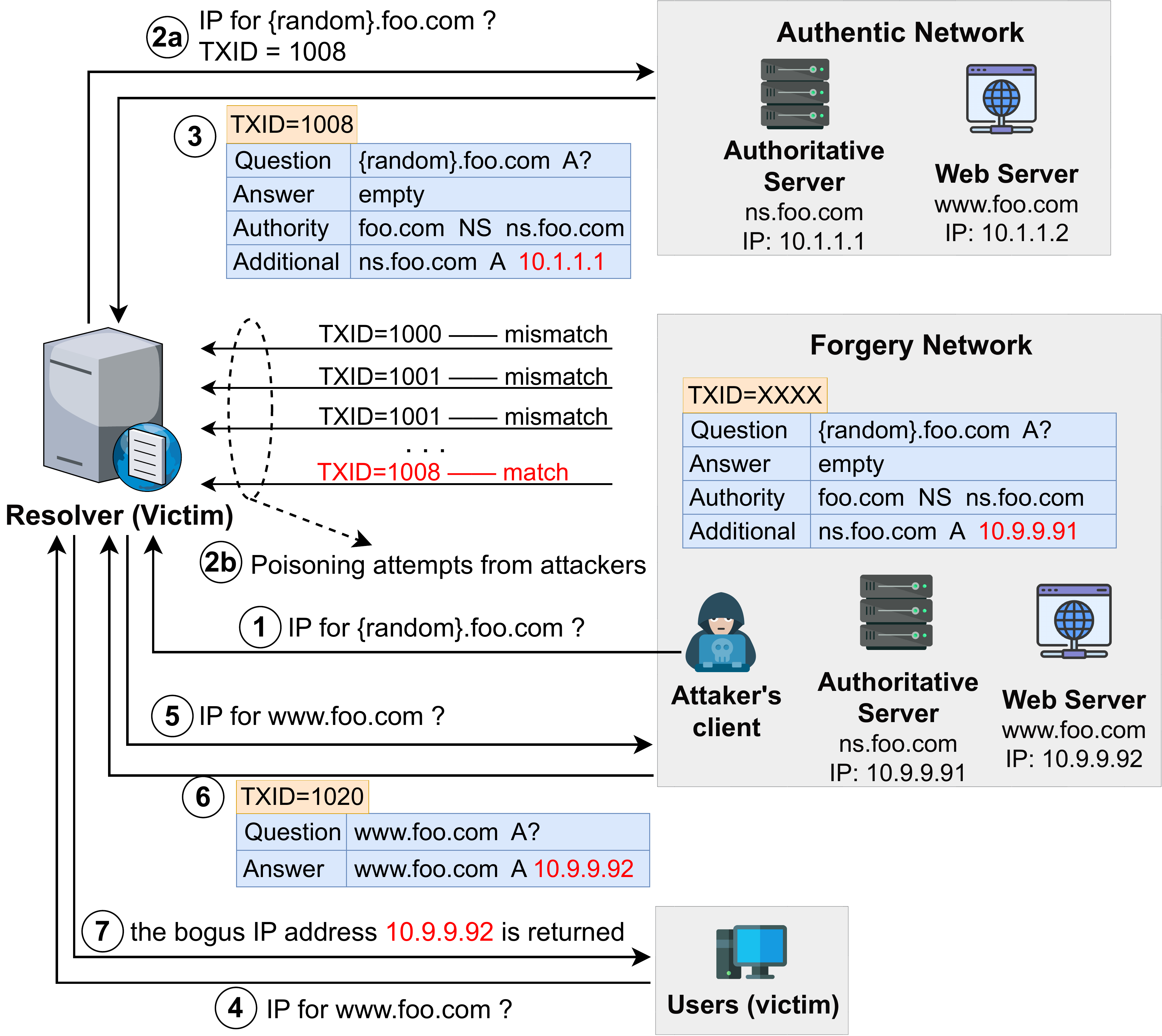}
\caption{Diagram of Kaminsky DNS cache poisoning.}
\label{fig-cache-poisoning}
\end{figure}

Depending on the capability of attackers, DNS cache poisoning can be categorized into three types: \textit{on-path}, \textit{in-path}, and \textit{off-path}\cite{chau2018adaptive}. On-path attackers can observe and modify network traffic, making it simple for them to intercept DNS requests/responses and tamper with contents. In-path attackers can impose their own device or software on the data transmission path. However, these attacks typically occur when the attacker controls the entire network traffic (e.g., powerful nation-state adversaries), thus often used in implementing censorship\cite{klein2017internet, duan2012hold}. Off-path refers to attackers not being directly positioned on the data transmission path but still able to send forged responses. Such an attacker leverages the predictable nature of transaction ID (TXID) and the limited TXID space in the DNS query, making their forged responses reach the target resolver before authentic ones. 

One notable instance of off-path cache poisoning was discovered and disclosed by Dan Kaminsky \cite{kaminsky2008black}. As an attack that can be launched continuously, it reduces the attack time from weeks to seconds and has attracted widespread attention from academics over the last decade\cite{zhang2021study}. In this paper, our analysis and experiments mainly focus on Kaminsky cache poisoning. However, in theory, our technique is applicable to all DNS cache poisoning scenarios.

Fig.~\ref{fig-cache-poisoning} shows the diagram of Kaminsky DNS cache poisoning, the key steps include:

\begin{itemize}
    \item \textbf{Step 1.} The attacker sends a query to the victim resolver for a random name (\textit{\{random\}.foo.com}), something unlikely to be in the cache of the resolver. 
    
    \item \textbf{Step 2a.} The resolver then queries the authoritative server, which may require multiple iterative queries.
    
    \item \textbf{Step 2b.} Simultaneously with step 2a, the attacker generates a list of random TXIDs and starts flooding the resolver with forged packets to guess the TXID issued in step 2a. The forged packets contain the NS record for \textit{foo.com} associated with a glue record.
    
    \item \textbf{Step 3.} The authentic server provides a legitimate response. However, if the attacker successfully guesses the TXID in step 2b, this authentic reply will be ignored since it arrives too late.
    
\end{itemize}

The attacking phase is now over, and the resolver has cached the poisoned record. The next time a user visits the website of \textit{www.foo.com} (\textbf{step 4}), the resolver queries the fake authoritative server referred by the forged glue record (\textbf{step 5}). The fake server returns an A record pointing to a malicious website (\textbf{step 6}). This record is also cached by the resolver and returned to the user (\textbf{step 7}). 


\subsection{Defense Solutions to DNS Cache Poisoning}

Significant research efforts have been devoted to developing effective countermeasures against DNS cache poisoning. These methods can be broadly categorized into three groups: 

\textbf{Challenge-response Defenses.} In a traditional DNS request, the transaction ID (TXID) space (i.e., the challenge word) is relatively small with only $2^{16}$ possibilities, making it possible to guess the TXID and forge the response. The challenge-response defenses aim to introduce randomization and entropy into the DNS request/response. Source Port Randomization (SPR)\cite{rfc5452} was first proposed and began to be adopted by several major DNS implementations, introducing randomness into the source port of the message field. However, SPR was discovered to have many vulnerabilities. A recent research\cite{man2020dns} showed how to use side channels to guess the UDP source port used by the resolver. 
Another challenge-response proposal is 0x20 encoding\cite{dagon2008increased}, which involves randomly capitalizing letters in the domain's query. This proposal is simple and effective, especially for lengthy names. However, in practice, it has severe compatibility issues with the authoritative servers on the Internet. A recent study\cite{zhang2021study} found that only one (i.e., OneDNS) of the 14 well-known public DNS resolvers supports this proposal. Additionally, randomizing the choice of authoritative servers\cite{herzberg2013fragmentation} is another proposal, which increases entropy depending on the number of authoritative servers. Unfortunately, it has been discovered that an attacker can cause query failures towards specific authoritative servers, which successfully binds a resolver to the one remaining server\cite{herzberg2012security}.

\textbf{Cryptographic Defenses.} 
Cryptographic-based methods provide another line of defense. DNSSEC (DNS Security Extensions)\cite{rfc6840} is a prominent solution that employs digital signatures to ensure the authenticity and integrity of DNS data. However, the success of DNSSEC requires cooperation and implementation by both the authoritative servers and resolvers throughout the DNS hierarchy. According to the measurement data from AP-NIC, the current DNSSEC deployment rate is just over 29.5\%\cite{APDNSSEC}. Besides, another data\cite{TLD2023} revealed that just 4.1\%, 4.8\%, and 5.3\% of .com, .net , and .org domains are signed. Another big challenge is that DNSSEC introduces additional overhead in computational resources and network bandwidth caused by the added cryptographic signatures\cite{lu2019end}. To alleviate this issue, a lightwave solution called ODD (On-Demand Defense)\cite{wang2018demand} was proposed, which greatly lowers the overhead at the expense of a slight security degradation. However, it can only be served as a transition solution for speeding DNSSEC adoption due to its security vulnerabilities.

Other cryptography-based approaches such as DNS-over-TLS (DoT)\cite{rfc7858}, DNS-over-HTTPS (DoH)\cite{rfc8484}, and DNSCrypt\cite{denis2015dnscrypt} have also been proposed primarily to protect user privacy and prevent MitM attacks. Although they have been standardized and are gaining strong support from the industry, issues such as misconfigurations were spotted on many services in practice\cite{lu2019end}. More efforts should be made to promote its correct and widespread adoption.

\textbf{Redundant  Architecture.} 
Another approach to mitigate cache poisoning involves the deployment of redundant architecture, which is typically built on top of a P2P (peer-to-peer) network and uses multiple resolvers to increase tolerance. In DoX\cite{yuan2006dox}, resolvers are linked to each other and share the information of verified records. HARD-DNS\cite{gutierrez2010hard} employs the quorum technique and receives responses from multiple resolvers to obtain a reliable one. DepenDNS\cite{alfardan2011analysis} designs a matching algorithm to choose the dependable IP addresses from multiple resolvers as output. However, although these solutions can solve external off-path cache poisoning attacks well, they are all vulnerable to internal attacks (especially Byzantine malicious attacks). Moreover, when the queries are multiplied, their resolution latency will also increase significantly, making their adoption unlikely.

\subsection{Blockchain-based DNS Decentralization Solutions}

In recent years, researchers have started exploring the application of blockchain technology to enhance DNS security.

\textbf{Blockchain and Smart Contract.} Blockchain is a decentralized digital ledger that records transactions sent by users, ensuring transparency, security, and immutability. Blockchain was first introduced by Bitcoin in 2008\cite{nakamoto2008bitcoin}, which is used for cryptocurrency. Later, the emergence of smart contracts\cite{androulaki2018hyperledger} promoted its broader application.
Smart contracts are self-executing contracts with predefined rules and conditions encoded in code, which is verifiable and tamper-proof. Users can interact with contracts by sending blockchain transactions.

\textbf{Application in DNS.} Namecoin\cite{loibl2014namecoin} is one of the earliest blockchain-based projects for decentralized domain registration and resolution under .bit top-level domain. Blockstack\cite{ali2016blockstack} is an improvement of Namecoin, which aims to build a user-centric internet by combining off-chain data storage. However, both of them are developed on Bitcoin. Bitcoin's slow transaction speed and lengthy confirmation time are inherited. Besides, these revolutionary architectures are incompatible with current DNS, making their adoption doubtful.

Some recent sultions began to utilize permissioned blockchains, making them more practical and compatible. TD-Root\cite{he2020td} presents a decentralized DNS root management architecture, removing the security vulnerabilities and trust risks in current centralized root management. RootChain\cite{zhang2021blockchain} goes further and decentralizes TLD (Top-Level Domain) data publication by empowering delegated TLD authorities to publish authenticated data. DNSonChain\cite{jin2021dnsonchain} proposes a blockchain-based naming service, which allows domain owners to claim their domain ownership. However, all these solutions manage to change DNS by concentrating on its authoritative side. And they do not specifically address the issue of cache poisoning. To our knowledge, TI-DNS is the first blockchain-based approach to counter cache poisoning attacks on the DNS resolver's side.

%% file: sys.tex
\begin{figure*}[tb]
\centering
\includegraphics[width=0.8\textwidth]{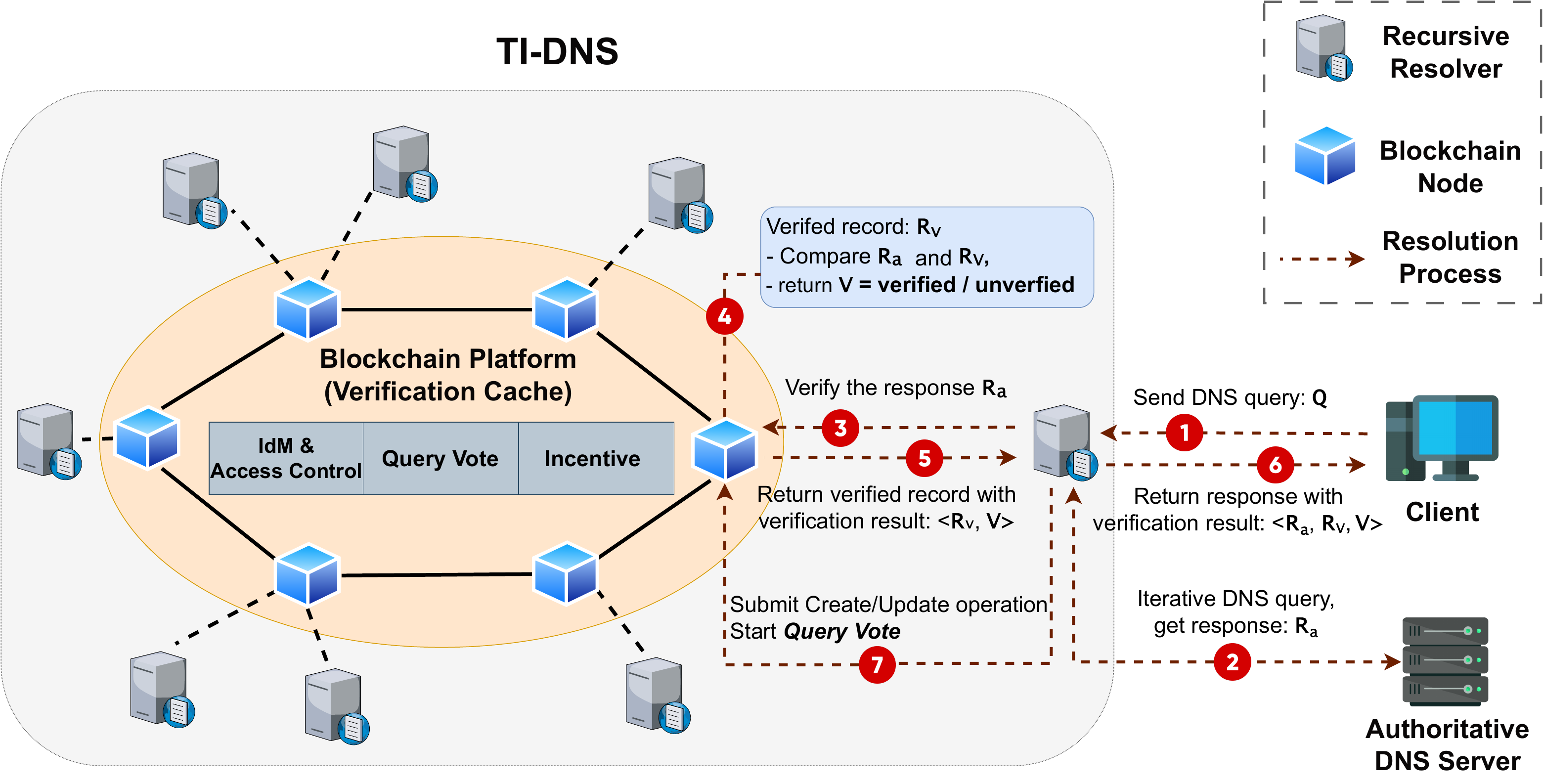}
\caption{High-level system architecture and resolution process of TI-DNS.}
\label{fig-architecture}
\end{figure*}

\section{System Design}
\label{sec-design}

TI-DNS’s core function is to verify the authoritative DNS responses received by resolvers and store the verified DNS records in the \textit{Verification Cache}. Compared to the traditional DNS resolution process, TI-DNS adds the step of verifying the authoritative response by querying the \textit{Verification Cache}. The \textit{Verification Cache} is stored on the blockchain platform as an immutable ledger, and is shared and maintained by all participants (i.e., resolvers). 
As shown in Fig.~\ref{fig-architecture}, there are two main components of TI-DNS:

\textbf{Recursive Resolver} is the entry point for DNS resolution. It acts as a middleware between clients and authoritative servers. In TI-DNS, after receiving a response $R_a$ from an authoritative server, the resolver will verify the response record by querying the \textit{Verification Cache}. Subsequently, the resolver responds $R_a$ and the verification result $V$ to the client. If the result is ``unverified'', the previously verified record $R_v$ will also be included in the response. Meanwhile, the resolver will submit the new record $R_a$ by invoking the create or update operation of the smart contract, thus initiating the subsequent \textit{Query Vote} process to verify its authenticity.

\textbf{Blockchain Platform} contains a distributed ledger that functions as a \textit{Verification Cache}  for resolvers. To guarantee the credibility of ledger, TI-DNS provides a \textit{Query Vote} mechanism, in which multiple resolvers are chosen as validators to vote on each \textit{Verification Cache} create or update operation. In addition, to incentivize participation and prevent malicious behavior, TI-DNS rewards resolvers that successfully provide valid records and validators that act appropriately. The aforementioned \textit{Query Vote} and \textit{Incentive} mechanisms are implemented in the form of smart contracts. Besides, the blockchain will assign a unique identity to each participant (i.e., resolver) and perform access control.

\subsection{Distributed Ledger Data Model}

Distributed ledger is a fundamental concept in blockchain platforms that stores crucial information about objects; both their current attribute values and the history of the transactions that led to those values. At each blockchain node, ledgers are stored as key-value pairs in a database. 
In TI-DNS, we specify data models for three separate ledgers:

\begin{figure}[tb]
\centering
\includegraphics[width=1\columnwidth]{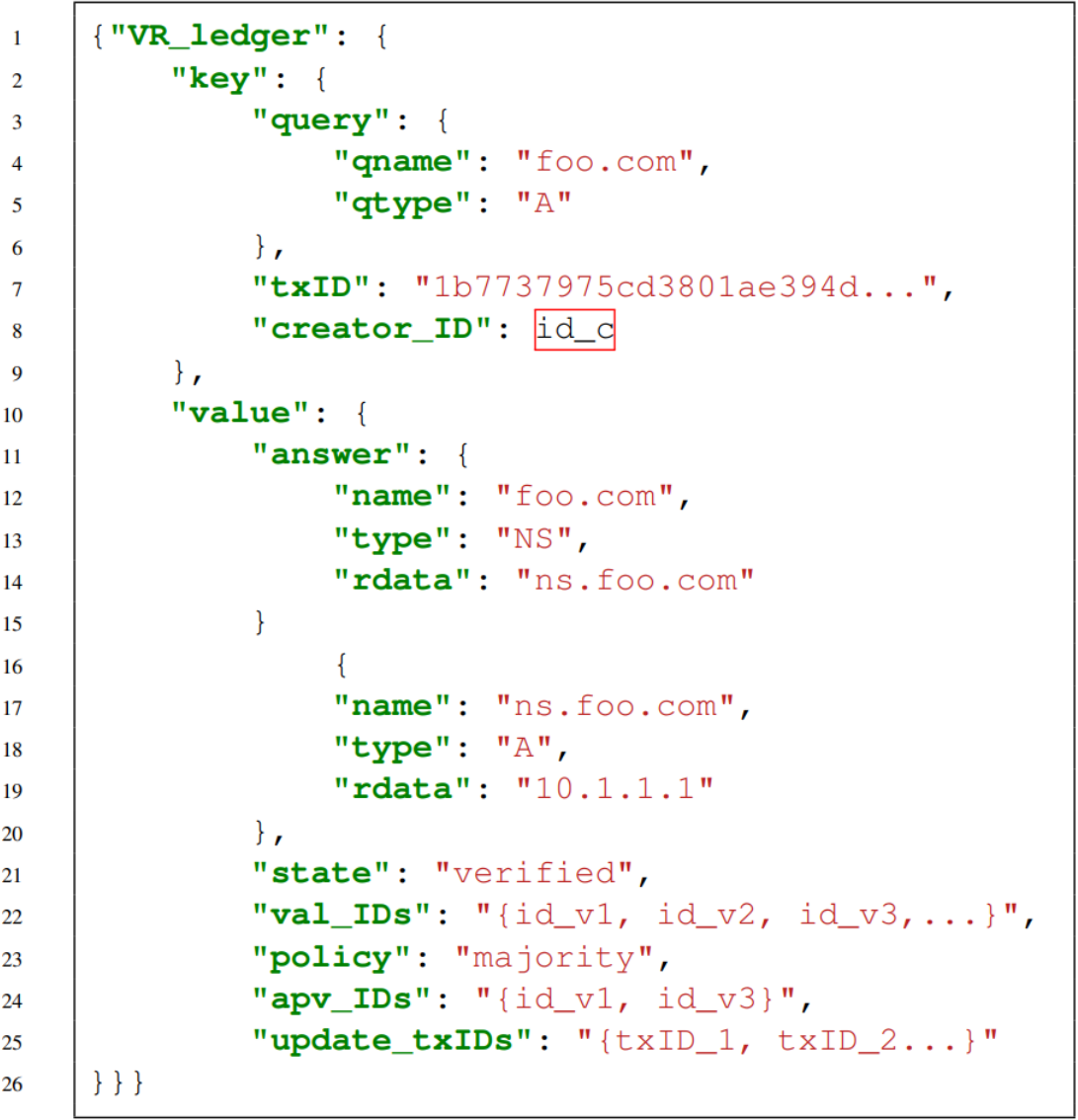}
\caption{A state of the \textit{VR\_ledger} in JSON format.} 
\label{list1}
\end{figure}

\textbf{VR\_ledger} (Fig.~\ref{list1}) stores previously verified records and is similar to a regular DNS cache, except it has no TTL restriction and contains additional attributes to assist \textit{Query Vote}. The \textit{key} in VR\_ledger consists of three fields: \textit{query} indicates the DNS query content, including \textit{qname} and \textit{qtype}; \textit{txID} and \textit{creator\_ID} respectively serve as a unique identifier of the blockchain transaction and identity that creates this record. The \textit{value} contains five fields: \textit{answer} records the verified answers; \textit{state} represents the verification state, which can be ``verified'' or ``unverified''; \textit{update\_txIDs} records a set of transaction IDs, which point to records that may need to be updated; \textit{val\_IDs} stores the identity set of the validators (i.e., voters); \textit{policy} determines the minimum number of approval votes necessary to pass the \textit{Query Vote}; \textit{apv\_IDs} stores the identity set of the voters who finally cast an approval vote. 

\textbf{Vote\_ledger} (Fig.~\ref{list2}) records the votes of the validators. The \textit{key} consists of two fields: \textit{voter\_ID} holds the identity of the voter; \textit{VR\_txID} stores the transaction ID of the record that needs to be validated in the VR\_ledger. The \textit{value} field stores the voting \textit{result}, which can be either ``yes'' or ``no''.

\begin{figure}[tb]
\centering
\includegraphics[width=1\columnwidth]{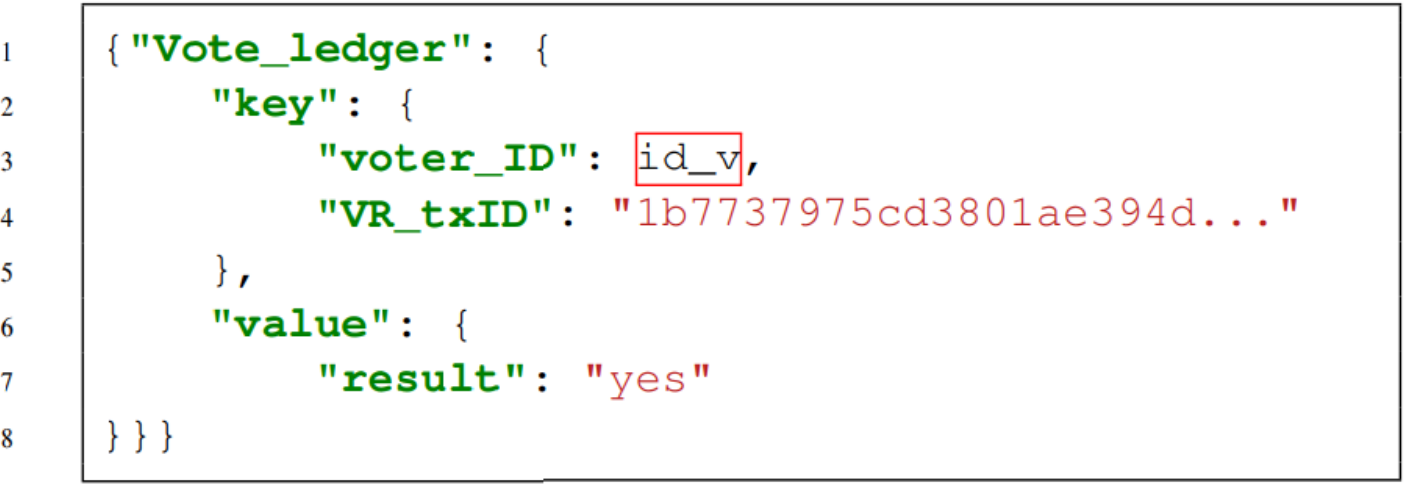}
\caption{A state of the \textit{Vote\_ledger} in JSON format.} 
\label{list2}
\end{figure}

\textbf{TokenOP\_ledger} (Fig.~\ref{list3}) records the updated amount of stakes held by each resolver. The \textit{key} contains two fields: \textit{ID} stores the identity of the resolver; \textit{tmp} corresponds to the timestamp of the stake update transaction. The \textit{value} field records the updated stake amount. In particular, rather than attempting to update a single row with the total amount of stake of each resolver, we design the TokenOP\_ledger to accept each transaction as an additive delta to the stake. The reason for this design is that, when multiple transactions arrive simultaneously, there is a time gap between when a transaction is simulated on the peer (i.e., read-set is created) and when it is ready to be committed to the ledger. During this time, another transaction may have already modified the same value, resulting in a failure of the transaction. To solve this issue, the frequently updated value (i.e., stakes of the resolver) is instead stored as a series of deltas in our design.

\begin{figure}[tb]
\centering
\includegraphics[width=1\columnwidth]{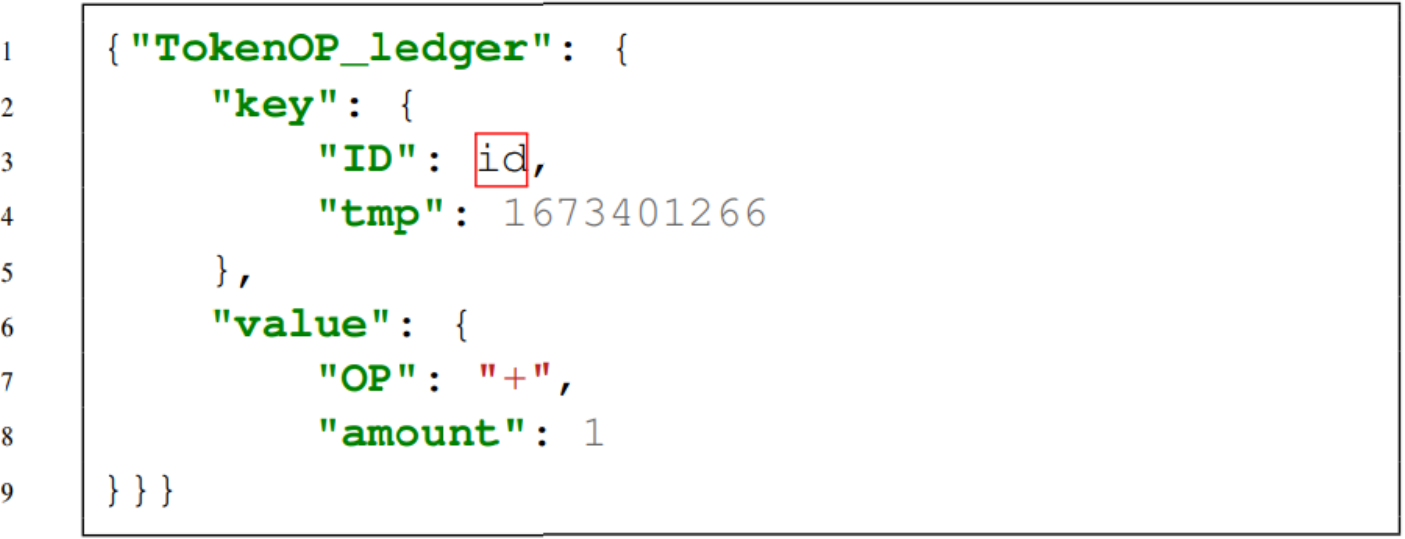}
\caption{A state of the \textit{TokenOP\_ledger} in JSON format.} 
\label{list2}
\end{figure}

\subsection{Resolution Verification}
\label{sec_Resolution Verification}

TI-DNS helps the resolver verify the authoritative responses during the DNS resolution. Fig.~\ref{fig-architecture} shows the above process:
\label{sec.ⅢC.1}

\begin{itemize}
    \item \textbf{Step 1.} The client initiates the DNS resolution process by sending a DNS query $Q$ to a resolver. Then, the resolver checks if the local cache contains verified records for the queried domain. If so, return the response directly.
    \item \textbf{Step 2.} The resolver iteratively queries the authoritative server to obtain the authoritative record $R_a$. However, as stated in Section~\ref{sec-kaminsky}, attackers may launch an attack at this step to forge the authoritative response.
    \item \textbf{Step 3-5.} The resolver searches the blockchain ledger for verified records $R_v$ and compares it with $R_a$ to obtain the verification result $V$, which is depicted in Algorithm~\ref{RV} and will be covered in more detail later.
    \item \textbf{Step 6.} The resolver responds $R_a$ to the client with the verification result $V$. If $V$ equals ``verified'', the resolver will cache $R_a$ to speed up the following resolution. Otherwise, the resolver will also include the previously verified record $R_v$ to the response.
    \item \textbf{Step 7.}  If $V$ equals ``unverified'', the resolver invokes the smart contract to submit the create or update transaction and starts \textit{Query Vote}, illustrated in Section~\ref{sec-QV}.

\end{itemize}

 In our pseudo-codes, interactions with blockchain are through either \textit{PutState} or \textit{GetState} function provided by built-in blockchain APIs. In addition, to support rich queries, functions about composite keys are also provided. \textit{CreateComKey} combines the given attributes together to form a unique composite key, and \textit{GetStaByParComKey} queries the state in the ledger based on the given partial composite key.

\begin{algorithm}
    \small
    \setstretch{1.05} 
    
    \SetKw{Init}{Initialisation:}
	\caption{Record Verification}
	\label{RV}
	\LinesNumbered 
	\KwIn{\textbf{QueryName}: $qname$, \textbf{QueryType}: $qtype$, \textbf{AuthoritativeRecord}: $R_a$}
	\KwOut{$out$}
	\Init{$update \leftarrow null$}
	\BlankLine 
	$iterator$ $\leftarrow$ GetStaByParComKey($VR\_ledger, qname, qtype$)\\
	$resid \leftarrow$ GetSubmittingClientIdentity()\\
	\While{$iterator$.HasNext()}{
	    $entry \leftarrow iterator$.Next()\\
	    $R_v \leftarrow$ GetAnswer($entry.value$)\\
	    $result \leftarrow$ CompareRecord($R_a, R_v$)\\
	    \If{$result = $ ``verified''}{
	    $out \leftarrow$ ($result, null$) \\
	    \Return $out$\\
	    }
	    $creator \leftarrow$ GetCreatorID($entry.key$)\\
	    $V_r \leftarrow$ GetValIDs($entry.value$)\\
	    \If{(($resid = creator$) $\vee$ ($resid \in V_r$))}{
	    $txid \leftarrow$ GetTxID($entry.key$)\\
	    $update$.Append($txid $)
	    }
	}
	$out \leftarrow$ ($result, update$) \\
	\Return $out$\\
\end{algorithm}
 
 \textbf{Record Vertification} function in Algorithm~\ref{RV} verifies the authoritative record $R_a$ received by the resolver. It begins by retrieving a record $iterator$ from the $VR\_ledger$ based on the provided $qname, qtype$ (line 2) and obtaining the submitting resolver's identity $resid$ (line 3). Due to the existence of one-to-many mapping between domains and IPs (e.g., CDNs use DNS-based redirection), verified records with the same query content but different answers are allowed to exist in the $VR\_ledger$. 
 Thus, the algorithm loops through each matching verified record $R_v$ (lines 4-6) and compares it with $R_a$ through the \textit{CompareRecord} function (line 7). For a domain with multiple addresses \textit{$IP_1$, $IP_2$, $IP_3$} mapped to it, one resolver may receive \textit{$IP_2$, $IP_1$, $IP_3$} while another might receive \textit{$IP_3$, $IP_2$, $IP_1$}. Thus we use \textit{set} comparison in the \textit{CompareRecord} function. If the comparison $result$ is ``verified'', we consider the verification success and return $result$ (lines 8-10). 
 Otherwise, the algorithm checks if the submitting resolver $res$ is the $creator$ or one of the validators $V_r$ of $R_v$ (line 13). In such cases, it retrieves the transaction ID $txid$ and appends it to the $update$ list for the following update operation (lines 14-15).
 After iterating through all matching records, the algorithm returns the verification $result$ associated with $update$ information (lines 16-17). Indeed, if we do not find any matching record in $VR\_ledger$, the algorithm will skip directly to the final step and return an ``unverified'' $result$.

\subsection{Query Vote}
\label{sec-QV}

\begin{algorithm}
    \small
    \setstretch{1.1} 
    
	\caption{Query Vote}
	\label{QV}
	\LinesNumbered 
	\KwIn{\textbf{Query}: $Q$, \textbf{AuthoritativeRecord}: $R_a$, \textbf{VotersNum}: $n$, \textbf{Policy}: $policy$, \textbf{UpdateTxIDs}: $update$}
	
	\BlankLine 
	\tcp{\small For \textbf{creator}:  \textit{Create/Update} transaction}
	Get $txid$, $creatorid$ through blockchain APIs\\
	$voters$ $\leftarrow$ VoterSelection($n$)\\
	$key$ $\leftarrow$ CreateComKey($Q$, $txid$, $creatorid$)\\
	$value$ $\leftarrow$ (\{\textit{answer} = $R_a$, \textit{state} = ``unverified'', \textit{val\_IDs} = $voters$, \textit{policy} = $policy$, \textit{update\_txIDs} = $update$\})\\
	PutState($VR\_ledger$).Append($key$, $value$)\\
	Send event message ($Q, R_a, txid$) to $voters$\\
	
    \BlankLine
	\tcp{\small For each \textbf{voter}: \textit{Voting} transaction}
	Get $voterid$ through blockchain APIs\\
	$R_d$ $\leftarrow$ DNS\_Lookup($Q$)\\
	$result \leftarrow$ CompareRecord($R_a, R_d$)\\
	$key$ $\leftarrow$ CreateComKey($voterid$, $txid$)\\
	PutState($Voter\_ledger$).Append($key$, $result$)\\
	Send event message $(txid, voterid, result)$ to $creator$\\
	
    \BlankLine
	\tcp{\small For \textbf{creator}: \textit{FinishVal} transaction}
	$entry$ $\leftarrow$ \rm GetStaByParComKey($VR\_ledger$, $Q$, $txid$)\\ 
	Obtain the event message to collect \textit{voting}\\
	Store the $voterid$ of \textit{voting} respectively in $apv$ and $disapv$ \\
	\If{$apv\_num$ \textgreater \ $threshold\_apv$}
	{
	    \If {$update \neq null$}
	    {
	        Delete the ledger state in $update$\\
	    }
	\rm PutState($VR\_ledger$).Update($entry.key$, \{\textit{state} = ``verified'', \textit{apv\_IDs} = $apv$\})\\
	Add stakes for $creator$ and approving voters $apv$\\ 
	}\Else{
	    Delete the ledger state of $txid$\\ 
	    Add stakes for disapproving voters $disapv$
	}

\end{algorithm}

To guarantee the authenticity of the verified records on the blockchain ledger, we propose a \textit{Query Vote} mechanism that selects multiple participants (i.e., voters) to validate each record create or update operation. As shown in Algorithm~\ref{QV}, we divide the process into three phases. The whole process is implemented through the smart contracts to ensure the accountability and transparency of the operation.

\textbf{Phase 1.} After receiving an ``unverified'' result in Algorithm~\ref{RV}, the resolver (i.e., creator) invokes the smart contract to create the new record $R_a$. The contract first selects a subset of $voters$ to serve as validators using an function described in Algorithm~\ref{VS}\ (line 2). Then, the creator appends $R_a$ into the $VR\_ledger$, with some additional attributes saved in $key$ and $value$, but the \textit{state} is ``unverified'' (lines 3-5). Finally, the contract constructs a validation request as $(Q, R_a, txid)$ and 
packages it into an event to notify $voters$ (line 6).
    
\textbf{Phase 2.} Upon receiving a validation request of $(Q, R_a, txid)$, each voter performs a voting operation. The voter sends a DNS query $Q$ to obtain the record $R_d$ from its local cache or the authoritative server (line 8). Then, the voter compares ($R_a, R_d$) to get the voting $result$ and appends it to $Vote\_ledger$ (lines 9-11). Finally, the voter creates an event as ($txid, voterid, result$) to notify the $creator$ (line 12).
    
\textbf{Phase 3.} The $creator$ listens to events emitted by $voters$ (line 14). If sufficient $voters$ agree with the validation request within a grace period, we consider the validation success and update the \textit{state} of the $VR\_ledger$ to ``verified'' (lines 16-19). To incentivize participation, the contract adds stakes for the $creator$ and approving voters $apv$ (line 20), elaborated in Section~\ref{sec.ⅢD}. Otherwise, we consider the operation a failure and delete the ledger state of the unverified record $R_a$ in the $VR\_ledger$ (lines 21, 22). Besides, the stakes of disapproving voters $disapv$ are increased (line 23).

\subsection{Voter Selection and Incentive mechanism} 
\label{sec.ⅢD}

TI-DNS selects a certain number of voters to act as validators in each voting process. The selection is made in a Proof-of-Stake pattern to defend against Sybil attacks, meaning the probability of each resolver being selected is proportional to its stakes. Besides, we assume that good voters own most stakes.

\begin{algorithm}
    \small
    \setstretch{1.05} 
    
    \SetKw{Init}{Initialisation:}
	\caption{Voter Selection}
	\label{VS}
	\LinesNumbered 
	\KwIn{\textbf{VotersNum}: $n$, \textbf{AllParticipants}: $participants$}
	\KwOut{$voters$}
	\Init{$total\_stake \gets 0$, $voters \gets []$}
	\BlankLine 
    \ForAll{$p \in participants$}
    {$total\_stake \gets total\_stake + p.stake$}
    \For{$i = 1$ to $n$}
    {
        $rand \gets \text{random number between 0 and }total\_stake$\\
        $cum\_stake \gets 0$\\
        \ForAll{$p \in participants$}
        {
            $cum\_stake \gets cum\_stake + p.stake$\\
            \If{$rand \le cum\_stake$}
            {
                $voters \gets voters + [p]$\\
                $participants \gets participants - [p]$\\
                $total\_stake \gets total\_stake - p.stake$\\
                \textbf{break}
            }
        }
    }
    \Return $voters$
\end{algorithm} 

\textbf{Selection Procedure.} As shown in Algorithm~\ref{VS}, the selection procedure starts by calculating the $total\_stake$ of all $participants$ (i.e., resolvers) in the system (lines 2, 3). For each voter to be selected, the algorithm generates a random number $rand$ between $0$ and the $total\_stake$ (line 5). It then iterates over the participants and accumulates their stakes until the cumulative stake $cum\_stake$ is greater than or equal to $rand$ (lines 6-9). The participant corresponding to this $cum\_stake$ is selected as a voter, and their stake is removed from the $total\_stake$ (lines 10-12). Finally, the algorithm returns the list of selected $voters$ (line 14).

\textbf{Stakes Incentive.} TI-DNS rewards resolvers for providing accurate records and voters that behave well. Specifically, once a voting process is completed, TI-DNS rewards voters whose votes align with the final voting result (i.e., only voters in the majority camp are rewarded). However, such an incentive is still not enough to prevent malicious behaviors. For example, a malicious participant may launch a Denial of Service (DoS) attack by sending a large number of transactions containing forged records, preventing legitimate transactions from being processed on time. To tackle this problem, each verified record create or update operation requires cost (i.e., stakes). When a resolver's stakes fall below a certain threshold, it is temporarily prohibited from submitting new records.

\section{Implementation and Evaluation}
\label{experiment}

We implement TI-DNS based on two open-source projects: 

 \textbf{Hyperledger Fabric} is a modular, permissioned blockchain framework that offers a flexible and scalable platform for building distributed ledger applications. Using a permissioned blockchain means that only authorized participants can access the network and perform transactions. We use Fabric to serve as the blockchian platform and develop smart contracts in Go.

\textbf{SDNS} is a fast and privacy-oriented DNS resolver implemented in Go, which supports many important features, including DNS caching, DNSSEC validation, DoH, and more. We utilize the plugin-based design of SDNS to customize and extend the functionality of the DNS resolver.

Based on the implementation, we evaluate the effectiveness and performance of TI-DNS by answering three questions: (1) how resistant is it to cache poisoning attacks? (2) what is the resolution performance of TI-DNS? and (3) what is the performance of blockchain write operations in TI-DNS? Meanwhile, we present how our design compares to (and outperforms) existing solutions.

\subsection{Experiment Setup}
We conduct experiments in a small network composed of 6 physical machines. Five are configured with 24 vCPUs of two Intel Xeon(R) Silver 4310 CPU at 2.1Hz and 32GB memory, while the other is allocated with 12 vCPUs and 32GB to serve as the client. The client machine is used to initiate DNS requests and smart contracts invocation requests to the network, and collects benchmark test results. All machines are connected to a local network with up to 1Gbps bandwith.

We use an open-source DNS performance tool dnsperf \cite{dnsperf} to evaluate the accurate latency and throughput of the resolver. Besides, to simulate the real DNS resolution traffic, we utilize Alexa's top domains\cite{alexa} to generate the DNS requests. For blockchain performance, we use the benchmark tool Hyperledger Caliper\cite{caliper}.

\subsection{Attack Resistance}
\label{sec-attack-experiment}

This test evaluates the ability of our system to defend against cache poisoning attack and compares it with several existing solutions: ODD\cite{wang2018demand}, HARD-DNS\cite{gutierrez2010hard}, and DepenDNS\cite{sun2009dependns}.

\begin{figure*}[tb]

\begin{minipage}[t]{0.32\linewidth}
    \centering
    \includegraphics[width=1\textwidth]{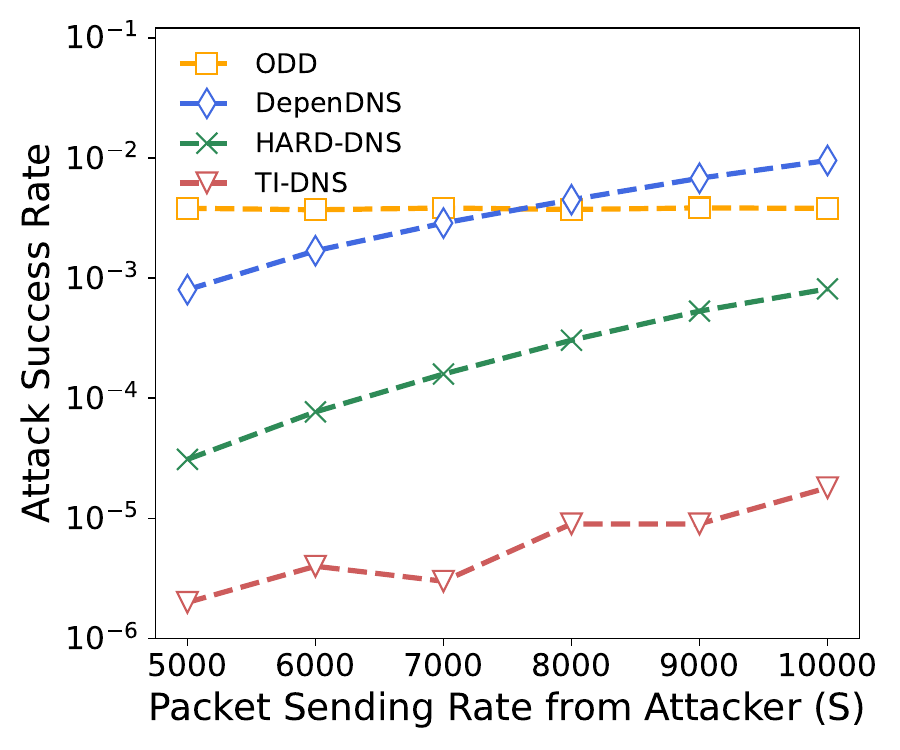}
    \caption{Comparison of attack resistance between TI-DNS and some existing solutions.}
    \label{fig-attack-all}
\end{minipage}%
\hfill
\begin{minipage}[t]{0.32\linewidth}
    \centering
    \includegraphics[width=1\textwidth]{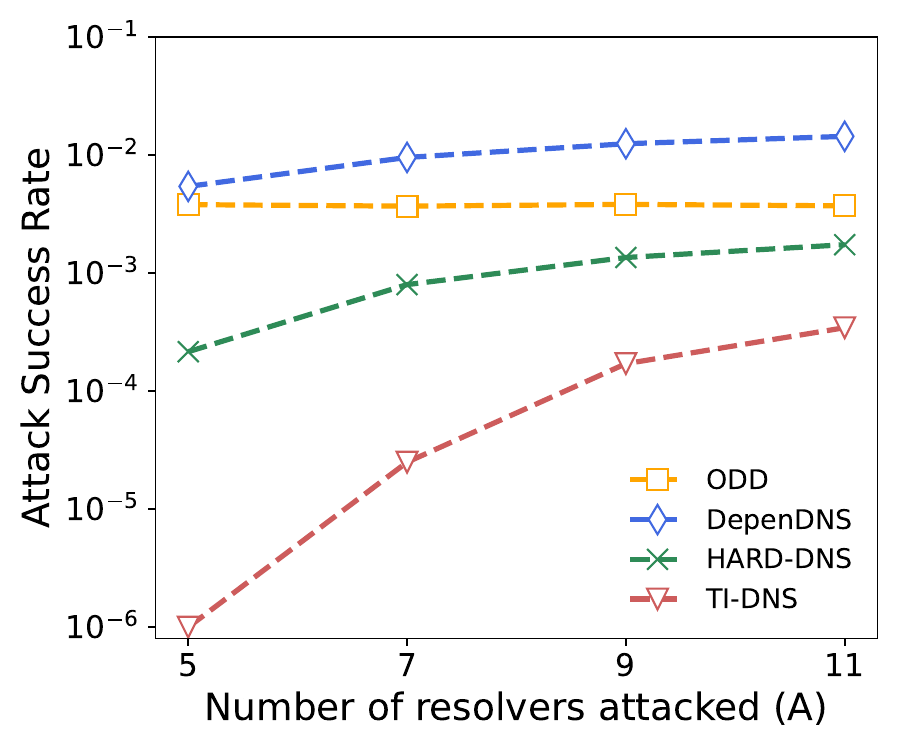}
    \caption{Attack success rate under different number of resolvers attacked ($N=12, V=9$).}
    \label{fig-attack-A}
\end{minipage}%
\hfill
\begin{minipage}[t]{0.32\linewidth}
    \centering
    \includegraphics[width=1\textwidth]{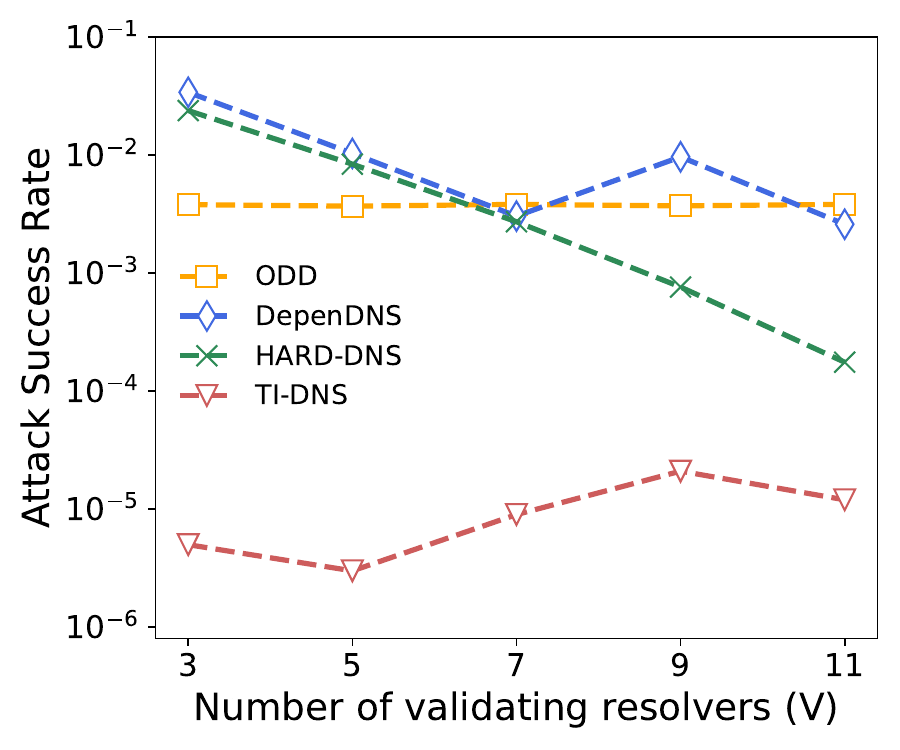}
    \caption{Attack success rate under different number of validating resolvers ($N=12, A=7$).}
    \label{fig-attack-V}
\end{minipage}%

\end{figure*}

As illustrated in Section~\ref{sec-kaminsky}, we simulate attackers exploiting multiple outstanding queries for the same question to maximize attack efficacy. The values of parameters are shown in Table~\ref{tab1}. Considering that the DNS queries we generate introduce excessive query load on authoritative servers for domains that we do not own, we build an authoritative server on the local network. Also, to simulate the time required for DNS resolution in a real environment, we manually set the response time $T_a$ of the authoritative servers to 0.2s, which is the window that attackers can utilize for each attack attempt.

\begin{table}[htbp]

\footnotesize
\centering
\renewcommand\arraystretch{1.3}
\caption{Parameters and their default settings}
\begin{center}
\setlength{\tabcolsep}{0.8mm}{
\begin{tabular}{|c|c|l|}
\hline

\textbf{Parameter} & \textbf{Value}& \makecell[c]{\textbf{Description}}  \\

\hline
$I$ & 65536 & Number of distinct TXIDs available     \\
\hline
$T_a$ & 0.2s & Response time of the authoritative server \\
\hline
$D$ & 50 & Number of identical outstanding queries \\
\hline
$S$ & $5\times10^3$qps & Packet sending rate form attackers \\
\hline
$N$ & 12 & Number of resolvers used \\
\hline
$V$ & 9 & Number of validating resolvers \\
\hline
$A$ & 7 & Number of resolvers attacked \\
\hline

\end{tabular}}
\label{tab1}
\end{center}
\end{table}

We conduct $10^6$ attack attempts for each round of experiment and calculate attack success rate. Fig.~\ref{fig-attack-all} shows the results under different packet sending rate from attackers ($S$). The attack success rate of TI-DNS is on the order of $10^{-6} \sim 10^{-5}$, while that of ODD and DepenDNS are $10^{-3}$, and HARD-DNS is $10^{-5} \sim 10^{-4}$. Besides, the probability of a successful attack grows as the packet sending rate from attackers ($S$) increases, except for ODD. In ODD, the resolver counts the incoming unmatched responses for each outstanding query. When the count amounts to a threshold of defense, the attacker traffic is identified, and the cryptographic defense mechanism is triggered. However, even in the worst-case scenario ($S = 10^4$ qps), the number of successful attacks on TI-DNS is just 18, far lower than the ODD (3804 attacks) and also lower than HARD-DNS (813 attacks) and DepenDNS (9516 attacks). 

We then further compare TI-DNS with similar redundant architectures (HARD-DNS and DepenDNS) under different settings where $S$ is fixed at $10^4$ qps. For consistency, we set their validation policy to ``majority''. Fig.~\ref{fig-attack-A} shows the result when we fix the number of validating resolvers ($V$) at 9 and change the number of resolvers attacked ($A$) from 5 to 11. The attack success rate of all methods is found to rise as $A$ increases, but TI-DNS has a more obvious advantage when A is small (if $A = 5$, TI-DNS has only 1 successful attack, while that of HARD-DNS and DepenDNS is 216 and 5437). A similar conclusion can also be drawn from Fig.~\ref{fig-attack-V}. If we fix $A$ at 7 and alter $V$ from 3 to 11. We see that other methods need a large $V$ to achieve the best resistance to attacks (11 for HARD-DNS and 7 for DepenDNS), whereas TI-DNS does not (only 5 successful attacks when $V = 3$).

The above phenomena is mainly due to the incentive mechanism in TI-DNS. For those few vulnerable resolvers or those with malicious behaviors, the proportion of stakes they own will decrease over time, as will their chances of being chosen as validators. On the contrary, honest resolvers who behave well will have a higher stake proportion. The \textit{Query Vote} process tends to choose such resolvers as validators in each round. Therefore, only a small $V$ is needed to achieve high attack resistance in TI-DNS. These features bring two significant advantages in practical applications: (1) In reality, attackers typically choose only one or several fixed resolvers as victims, in which case TI-DNS has the extremely high anti-attack ability; (2) TI-DNS does not require a large number of validators to participate in each \textit{Query Vote} procedure, which reduces the time consumption of this process.

\subsection{Resolution Performance}
\begin{figure}
    \centering
    \subfloat[Resolution Latency]
    {\includegraphics[width=0.5\columnwidth]{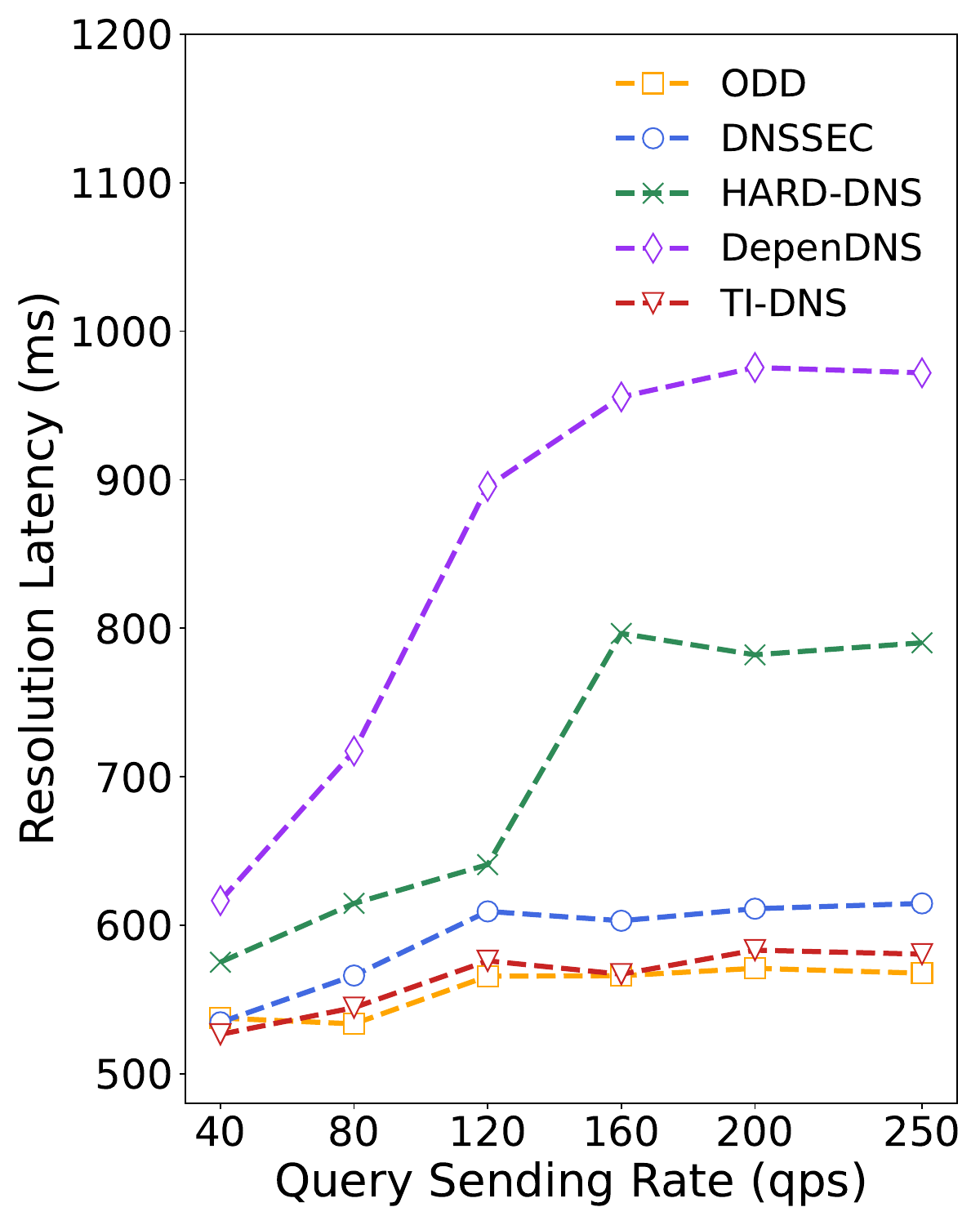}
    \label{fig_performace_latency}
    }
    \subfloat[Throughput]
    {\includegraphics[width=0.5\columnwidth]{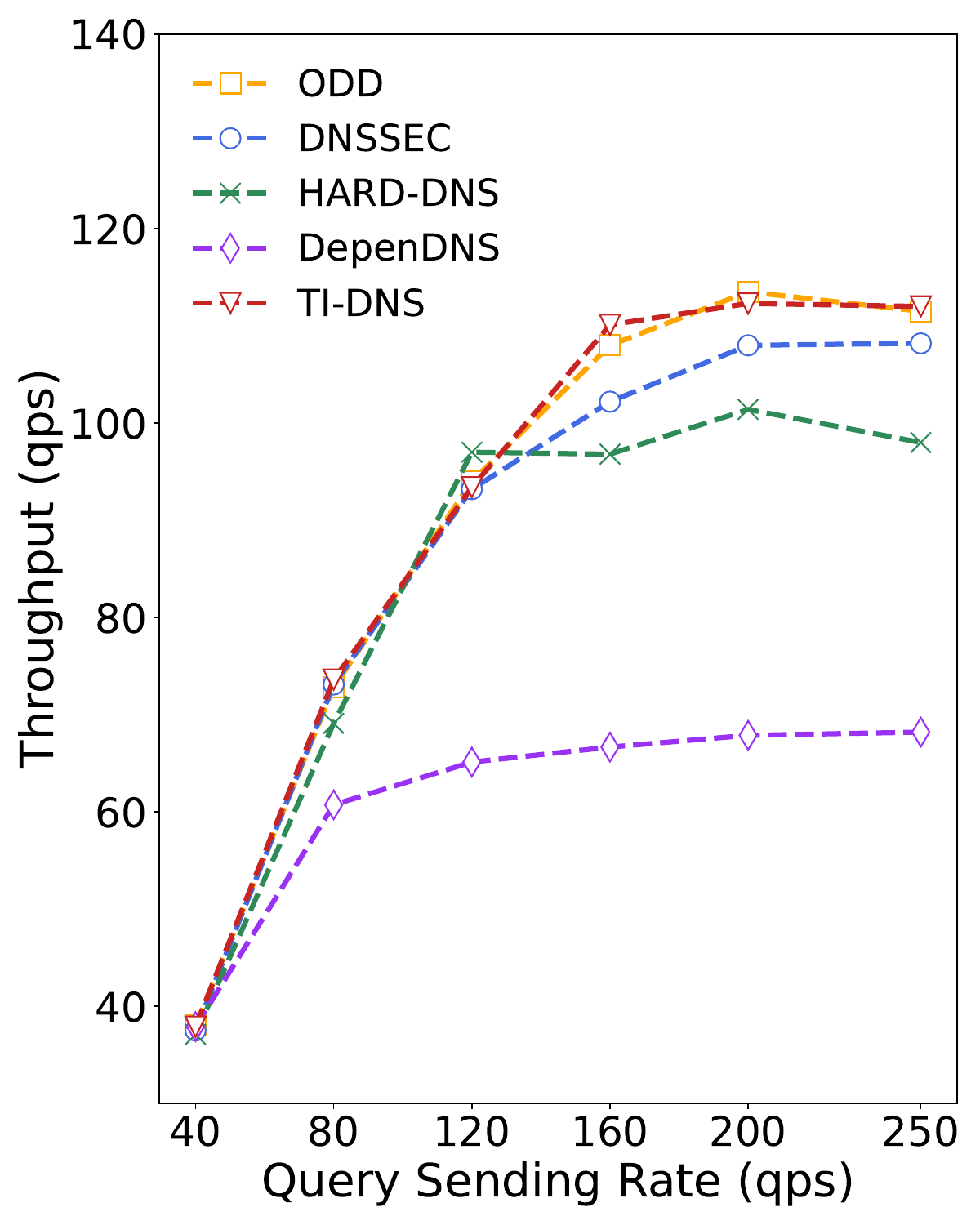}
    \label{fig_performace_throughput}
    }
    \caption{Comparison of resolution latency and throughput between TI-DNS and some existing solutions.}
    \label{fig_performace}
\end{figure}

We now analyze the resolution performance of TI-DNS. In this experiment, we construct DNS queries using top 10,000 Alexa's domains. In each round of simulation, the attacker attempts to launch an attack every 500 ms by sending forged packets. Fig.~\ref{fig_performace} shows the average resolution latency and throughput under different query sending rate. The result shows that as the query sending rate increases, the latency and the throughput increase. After passing the bottleneck point (around 160 qps), the latency and throughput stabilize. Unless otherwise stated, we analyze and compare the results when the sending rate is 200 qps in the following text.

Firstly, compared to the cryptographic-based DNSSEC method (611 ms, 108 qps), TI-DNS (583 ms, 113 qps) becomes 5\% faster, and the throughput is 4.6\% higher. This is due to the processing costs of implementing encryption and decryption at authoritative servers and resolvers in DNSSEC.
ODD significantly lowers the DNSSEC overhead, making its performance (572 ms, 115 qps) comparable to ours. However, if considering the worst-case scenario, that is, the attacker constantly sends the forged responses at a high rate towards the target resolver, ODD will switch to DNSSEC-aware mode. In this mode, its performance is identical to DNSSEC's. In TI-DNS, the resolution performance of the target resolver will not be affected even in the worst-case scenario.

Secondly, TI-DNS shows a significant advantage compared to other redundant architectures. Regarding latency, TI-DNS is 34\% faster than HARD-DNS (782 ms) and 67\% faster than DepenDNS (976 ms). In terms of throughput, TI-DNS is 11\% higher than HARD-DNS (101 qps) and 65\% higher than DepenDNS (68 qps). Their poor performance is because they require multiple resolvers to handle a single request, and as the number of resolvers increases, their resolution latency increases accordingly, but TI-DNS will not. In Fig.~\ref{fig_performace}, we set the number of resolvers used ($V$) as 5. If $V$ is set to 3, we find that DepenDNS's performance (688 ms, 108 qps) is still inferior to TI-DNS's (586 ms, 111 qps), whereas HARD-DNS becomes slightly better (552 ms, 122 qps). However, as shown in Fig.~\ref{fig-attack-V}, a too small value of $V$ will cause a great loss in the ability for HARD-DNS and DepenDNS to resist attacks.

In conclusion, TI-DNS achieves a strong ability to defend against cache poisoning attacks while sacrificing less resolution performance compared to other solutions, which enables its practical application.

\subsection{Blockchain Write Operations Performance}

In TI-DNS, when the resolver finds that the authoritative response does not match the verified record on the blockchain, it invokes the smart contract to start the \textit{Query Vote} of a new record. This process includes three different write operations (\textit{Create/Update Record}, \textit{Voting} and \textit{Finish Validation}) to the blockchain, each synchronized by consensus across blockchain nodes in the form of a transaction and then executed.

\begin{figure}[tb]

    \centering
    \includegraphics[width=0.85\columnwidth]{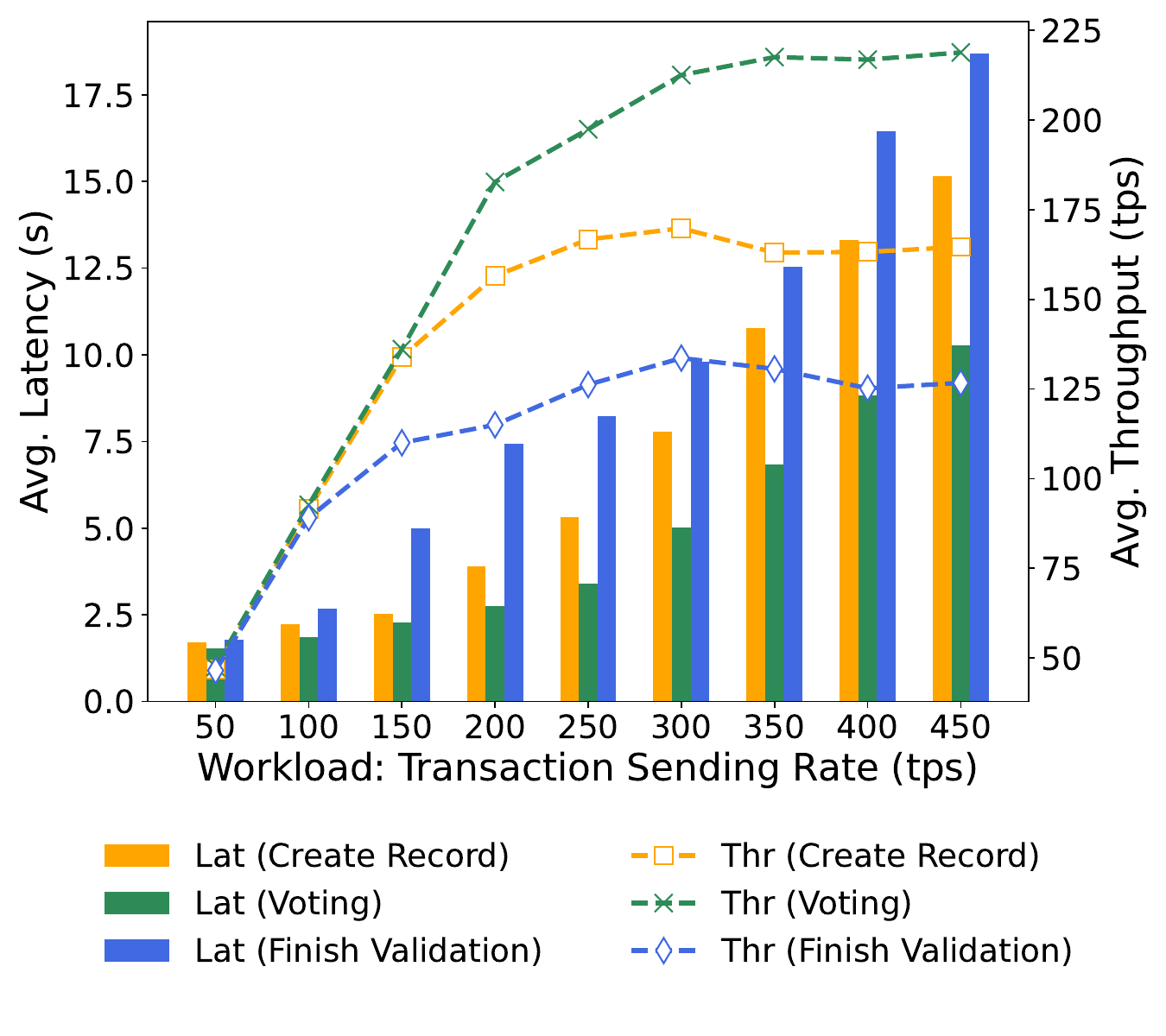}
    \caption{Average throughput and latency of TI-DNS's write operations on blockchain under different workloads.}
    \label{fig_Blockchain_throughput_and_latency_5o5p}

\end{figure}

\textbf{Write Operation Performance.} Fig.~\ref{fig_Blockchain_throughput_and_latency_5o5p} depicts the changing trend of the average latency and throughput of three separate operations under different workloads, from 50 tps to 450 tps. The blockchain network contains five committee members, each maintaining an orderer node and a peer node. The result shows that although the performance varies due to the complexity of different operations, the changing trends are similar. Therefore, unless otherwise stated, the following content uses the \textit{Create/Update Record} operation to display and analyze the results. At the end of this section, we will analyze the consumption of the entire \textit{Query Vote} process as a whole.

Fig.~\ref{fig_Blockchain_throughput_and_latency_5o5p} shows that the throughput starts to increase linearly with the growth of workload, peaking at 170 tps when the transaction sending rate hits 300 tps. However, while the workload keeps growing, the throughput begins to remain constant or perhaps decline slightly. Meanwhile, the latency also reflects a similar phenomenon. In the beginning, the latency increases slowly with the workload. When the sending rate reaches 250 tps, the latency is acceptable at 5.32s. However, as the workload further increases, the latency increases rapidly, reaching 7.78s at 300 tps (increased by 46\%) and 13.31s at 400 tps (increased by 150\%). The reason is that in Fabric, once the sending rate reaches a critical point (around 250-300 tps), the processing queue of the nodes begins to block, increasing the average waiting time and even the failure rate of transactions.

\textbf{Practically Analysis.} Fig.~\ref{fig_Blockchain_throughput_and_latency_5o5p} indicates that the bottleneck of the entire \textit{Query Vote} process lies in the \textit{Finish Validation} operation, while it can still reach 110 tps with 5 committee members (the sending rate is 150 tps). This means that TI-DNS can handle at least 110 requests per second for the domain to IP mapping changes, which far exceeds the change frequency of a regular domain. From the perspective of latency, the complete \textit{Query Vote} process has a latency of around 9.82s (2.53s for \textit{Create/Update} Record, 2.29s for \textit{Voting} and 5s for \textit{Finish Validation}). Since the writing operations of the blockchain and the DNS resolution are asynchronous, the latency is insensitive and acceptable to users. 

Furthermore, it should be emphasized that in TI-DNS, the number of peer nodes and resolvers are in a one-to-many relationship, that is, multiple resolvers can connect to the same peer using different identities. Besides, if the blockchain network needs to add more peers but cannot tolerate excessive performance loss, the network manager can still boost performance by sacrificing partial decentralization (e.g., reducing the number of orderer nodes, modifying endorsement policies).

\section{conclusion}
\label{conclusion}

In this paper, we proposed TI-DNS, a blockchain-based DNS resolution architecture, to detect and correct the forged DNS records caused by the cache poisoning attacks. TI-DNS includes a multi-resolver \textit{Query Vote} method and a stake-based \textit{Incentive} mechanism, guaranteeing the credibility of verified records on the blockchain ledger and encouraging well-behaved participation. Finally, we implemented the prototype of TI-DNS and conducted comprehensive experiments, demonstrating its effectiveness and feasibility.


\section*{Acknowledgment}

This work is funded by NSFC Grant No. 62202450, Huawei New IP open identification resolution system project No. TC20201119008 and Postdoctoral Exchange Program No. YJ20210185.